\documentstyle[12pt,titlepage]{article}

\advance\textheight by \topskip

\author{
  Hartmut Frommert\\
  \rm E-Mail: \tt Hartmut.Frommert@uni-konstanz.de\\[1ex]
  \em Dept.\ of Physics, University of Constance\\
  \em P.O.Box 55 60 M 678, D-78434 Konstanz, Germany
}

\title{\bf On the coupling of massless particles to scalar fields}

\newcommand{\beq}{\begin{equation}}
\newcommand{\eeq}{\end{equation}}
\newcommand{\bea}{\begin{eqnarray}}
\newcommand{\eea}{\end{eqnarray}}
\newcommand{\beas}{\begin{eqnarray*}}
\newcommand{\eeas}{\end{eqnarray*}}
\newcommand{\beqs}{\begin{displaymath}}  
\newcommand{\eeqs}{\end{displaymath}}    

\newcommand{\bm}[1]{{\mbox{\boldmath$#1$}}}

\newcommand{\calc}{{\cal C}}     
\newcommand{\cale}{{\cal E}}     \newcommand{\calf}{{\cal F}}

\newcommand{\calo}{{\cal O}}

\newcommand{\G}{{\rm G}}

\begin{document}

\maketitle

\thispagestyle{empty}
\section*{Abstract}
It is investigated if massless particles can couple to scalar fields
in a special relativistic theory with classical particles.
The only possible obvious theory which is invariant under
Lorentz transformations {\em and\/} reparametrization of the affine
parameter leads to trivial trajectories (straight lines) for the
massless case, and also the investigation of the massless {\em limit\/}
of the massive theory shows that there is no influence of the scalar
field on the limiting trajectories.

On the other hand, in contrast to this result, it is shown that massive
particles {\em are\/} influenced by the scalar field in this theory even
in the ultra-relativistic limit.


\clearpage
\pagenumbering{arabic}

\section{Introduction}

In the context of light deflection in gravitational fields, it has been
pointed out that within the Newtonian theory of gravity (i.e., gravity
as a scalar field), energy conservation arguments lead to effects on
light propagation (Soldner 1801).
This can be seen more fundamentally by considering
classical particles, massive or massless, which propagate with light
velocity $c$ through a scalar field acting on them in the absolute
Newtonian spacetime.

Going over to the special-relativistic Minkowskian spacetime, the situation
gets different, as in this case there occurs a fundamental difference
between massive and massless particles, with respect to their propagation.
Within this theoretical framework, one can see quite soon that it is no more
clear if massless particles can couple to scalar fields at all.

That such a coupling is impossible has been claimed occasionally, but
without rigid proof, even in some textbooks such as Greiner 1989.
It is also of interest with respect to particle physics, where the Higgs
mechanism is used to generate mass by a scalar field, implying that coupling
to the scalar Higgs field occurs only for massive particles (more acurately,
even to the masses of the particles; this interaction has been studied in
detail by Dehnen and Frommert 1990, 1991; Dehnen {\em et.al.\/} 1990).
However, on the field theoretical level, it is possible to add coupling to
massless particle fields by hand, provided only that the scalar field has a
trivial ground state, simply such an interaction was not found in nature
and, therefore, can be very weak at best.

In order to investigate if massless particles can couple to scalar fields
in a special relativistic theory with classical particles, it is most
convenient to find a Lagrange function (and thus a theory) which is invariant
under Lorentz transformations {\em and\/} reparametrization of the affine
parameter which replaces time in special relativistic theories%
\footnote{
  If it should turn out that it should be impossible to find such a Lagrange
  function, one had to select a set of equations of motion by hand, and to
  face all the problems which arise in such theories.
}.
As outlined in this work, there is only one possible obvious theory of
coupling relativistic particles to scalar fields
which matches these requirements.
This theory turns out to be trivial for the case of massless particles,
i.e.\ the trajectories of lightlike particles are straight lines.
Moreover, the investigation of the massless {\em limit\/} of the more
general theory shows that there is no influence of the scalar field on the
limiting trajectories.
This result reproduces the wellknown fact that because of the observed light
deviation in gravitational fields, gravity cannot be described correctly by
a special-relativistic scalar theory in flat Minkowskian spacetime.

On the other hand, one may suspect that the problem of massless particles may
be correlated in some way with the relativistic limit of the massive problem.
This limit is also investigated here.
In difference to the result above, it turns out that massive particles
are deflected by the scalar field in this theory, even in the
ultra-relativistic limit, i.e.\ $V \longrightarrow c$.

\section{Lagrange functions for relativistic particles}

According e.g.\ to Greiner 1989 (eq. (98)) or Lindner 1994, the
special relativistic Lagrange function for a {\em massive\/},
free classical particle (i.e., ``test body'') is given by%
\footnote{
  $m_0$: rest mass, $c$: light velocity, $t$: time,
  $(\ldots)^\cdot = {d\over dt}(\ldots)$, 
  $V^\mu$: 4-velocity ($V^\mu=\dot x^\mu$).
  Convention: $\eta_{\mu\nu} = \mbox{\rm diag}(+---)$
}
\beq
L = - m_0 c \sqrt{V_\mu V^\mu}
\eeq
which must be inserted into the action principle:
\beq
S = S(t_1, t_2) = \int_{t_1}^{t_2} L\>dt
\eeq
This action is invariant under both Lorentz or Poincar\'e transformations,
and re\-pa\-ra\-me\-tri\-za\-tion of the "parameter" $t$,
as $L dt$ may be rewritten
with the proper time $\tau$ as parameter, as $L dt=-m_0 d \tau$.

From this action follows via Hamilton's variation procedure the
(vector-valued) Euler-Lagrange equation:
\beq
{d\over dt}\left({1 \over \sqrt{V_\alpha V^\alpha}}V^\mu\right) = 0\ ,
\label{eq:freemotion}
\eeq
the equation of motion for a (relativistic) free particle. It may be
evaluated to
\beq
\ddot x^\mu
- {\dot x^\mu \dot x_\nu \over \dot x^\alpha \dot x_\alpha} \ddot x^\nu
= \left(\delta^\mu_\nu - 
        {\dot x^\mu \dot x_\nu \over \dot x^\alpha \dot x_\alpha}\right)
  \ddot x^\nu
= 0\ .
\eeq
It is remarkable that because of the projector, these are only 3 nontrivial
independent components of this equation (scalar equations):
Contraction with $V_\mu=\dot x_\mu$ yields identically zero.
Equation (\ref{eq:freemotion}) may also be contracted with
$u_\mu=V_\mu/\sqrt{V_\beta V^\beta}$ to enhance
$u^\mu u_\mu=const\in\{0,\pm1\}$.

One may straightforwardly add e.g.\ electromagnetism (as an external field)
by adding an interaction term in the following way%
\footnote{
  $A_\mu$: Electromagnetic 4-potential,
  $q$: Electromagnetic (i.e.\ electric) charge.
}:
\beq
L = - m_0 c \sqrt{V_\mu V^\mu} + q A_\mu V^\mu
\eeq
yielding the equations of motion for a massive particle in the
electromagnetic field:
\beq
{d\over dt}\left({m_0 c \over \sqrt{V_\alpha V^\alpha}}V^\mu\right) =
- q F^\mu_{\ \nu} V^\nu
\label{eq:em}
\eeq
with the usual electromagnetic field strength
$F_{\mu\nu} = A_{\nu,\mu} - A_{\mu,\nu}$.
This is, of course, the usual Lorentz force equation for a classical point
particle.
As $F_{\mu\nu}$ is skew symmetric, contraction with $V^\mu$ makes
{\em both\/} sides of (\ref{eq:em}) vanish identically, so we have again
only 3 independent nontrivial equations.

We remark here immediately that, if $A_\mu$ was the 4-gradient of a scalar
field (or that of some functional of arbitrary fields), say $\Phi({\bm x})$
the right-hand side would vanish identically because of the antisymmetry of
the field strength $F_{\mu\nu}$.
Therefore, the coupling of gradients to a particle via a potential energy
term in the Lagrange function,
\beqs
L_i = \Phi_{,\mu} V^\mu\ ,
\eeqs
does not contribute to the force on the particle. This may even be seen on a
more fundamental level: The term evaluates to a pure ``surface'' term in the
action:
\beqs
S_i = \int_{t_1}^{t_2} L_i \> dt
   = \int_{\calc;\bm x(t_1)}^{\bm x(t_2)} \Phi_{,\mu} dx^\mu
   = \Phi({\bm x(t_2)}) - \Phi({\bm x(t_1)})\ ,
\eeqs
which is constant if the standard convention of vanishing variations at
$t_1$ and $t_2$ is obeyed.
For this reason the coupling to scalar fields must be done in a different way.

For {\em massless\/} fields, the Lagrange function has to be modified:
The rest mass $m_0$ must be avoided, because it is identically zero.
Therefore, we replace the factor $m_0 c$, which is a ``rest momentum'', by
another, equivalent constant, namely $\hbar/\bar\lambda$, i.e.\ an inverse
characteristic wavelength up to a factor $\hbar$ (for massive particles,
the wavelength $\bar\lambda$ is the Compton wavelength).

In view of the reparametrization invariance of our theory,
which is required if we demand the incorporation of space and time
into spacetime, as usual in (special and general) relativity,
the coordinate time $t$ can no longer play a unique role in our theory%
\footnote{
  This demand is forcing for {\em massless\/} particles, as there is no
  natural invariant parameter for them, such as the proper time for massive
  particles.
}.
Therefore, the time $t$ in our action can be replaced by any continuous
and monotonic parameters $\lambda$ along the curve:
$t\longrightarrow\lambda(t)$.
Then the velocity $V^\mu=dx^\mu/dt$ is replaced by
\beqs
v^\mu = {dx^\mu\over d\lambda}
\eeqs
It will be convenient in the following to choose the parameter $\lambda$
in such a way that
\beq
d\lambda=\bar\lambda dt\ ,\ \
v^\mu = {dx^\mu\over d\lambda} = {V^\mu\over \bar\lambda}\ .
\eeq
Then the Lagrange function and the action take the form:
\beq
L = - \hbar\sqrt{v_\mu v^\mu}
  = - \hbar\sqrt{\eta_{\mu\nu} {dx^\mu\over d\lambda} {dx^\nu\over d\lambda}}
\ ;\ \
S = {1\over\bar\lambda} \int_{\lambda_1}^{\lambda_2} L d\lambda\ .
\eeq

\section{Coupling scalar fields to relativistic particles}

As the simplest possibility, i.e. adding to the Langrange function a term%
\beqs
L_i = q \Phi_{,\mu} V^\mu
\eeqs
fails as outlined above, we must look for other couplings. In order to
keep the invariance of the action with respect to (arbitrary)
reparametrization, the relation $L_i\sim 1/d\lambda$ must be kept.
This leads to a first ansatz
\beq
L_i = - \hbar f(\Phi,\Phi_{,\mu},\ldots) \sqrt{v_\mu v^\mu}
\eeq
and thus the total Lagrange function
\beq
L = - \hbar F(\Phi,\Phi_{,\mu},\ldots) \sqrt{v_\mu v^\mu}
\eeq
with $F=1+f$, or (see e.g. Misner, Thorne, Wheeler 1973):
\beq
L = - \hbar e^\phi \sqrt{v_\mu v^\mu}
\label{eq:L0}
\label{eq:L}
\eeq
with $\phi=\ln F$.
Since the scalar field was in no way fixed otherwise on this level,
we may now regard $\phi$ as our new scalar field, or scalar potential.
This will be justified later by the analogy of the equation of motion to the
Newtonian equation with a scalar potential.

Now we take a closer look on the Lagrange function for classical massless
particles.
The general one-particle Lagrange
function $L(x^\mu,v^\mu)$ may be expanded in powers of $v^\mu$:
\beq
L = l_0 + (l_1)_\mu v^\mu + (l_2)_{\mu\nu} v^\mu v^\nu + \ldots
  = \sum_{n=0}^\infty (l_n)_{\mu_1\ldots\mu_n} v^{\mu_1}\ldots v^{\mu_n}\ .
\label{eq:Ltayl}
\eeq
Reparametrization invariance then requires
\beq
(l_n)_{\mu_1\ldots\mu_n} =
  { (f_n)_{\mu_1\ldots\mu_n} (x^\mu)
    \over \left(v_\alpha v^\alpha\right)^{(n-1)/2} }\ .
\eeq
As masslessness corresponds to the condition $v_\mu v^\mu=0$,
one is led to the conclusion that the power series (\ref{eq:Ltayl})
must stop after $n=1$, because otherwise divergences in $L$ would appear
(in the form of zero denominator terms).
Then equation (\ref{eq:L}) is the most general Lagrange density possible for
a classical particle interacting with a scalar field%
\footnote{
  The arguments given here are certainly valid for straightforwardly formed
  Lagrange functions such as those discussed here.
  It is however difficult to find a general decision if it is possible to
  find, within special relativity, some strange coupling term as a
  counterexample or not.
  One might think e.g. of replacing the Minkowski metric under the root
  factor, $\sqrt{\eta_{\mu\nu}v^\mu v^\nu}$, by some
  $g_{\mu\nu}=\eta_{\mu\nu}+h_{\mu\nu}(\Phi)$.
  This however would lead at least to the limits if not out of Special
  Relativity.
}.

Variation of the Lagrange function (\ref{eq:L}) w.r.t.\ the particle
trajectory yields the vector valued equation of motion:
\beq
0 = e^\phi \sqrt{v_\lambda v^\lambda} \phi_{,\mu}
    - {d\over d\lambda}
      \left({e^\phi v_\mu\over\sqrt{v_\lambda v^\lambda} } \right)
\label{eq:eom}
\eeq
which can be rewritten as
\beq
0 = e^\phi {1 \over \left(v_\lambda v^\lambda\right)^{3/2}}
    \left( v_\kappa v^\kappa \delta^\nu_\mu - v_\mu v^\nu \right)
    \left( v_\rho v^\rho \phi_{,\nu} -
           \eta_{\nu\rho} {d^2 x^\rho\over d\lambda^2} \right)
\label{eq:0mot}
\eeq

\noindent
As the exponential factor is positive, the rest of the equation must vanish.
Now with $v^\mu$ a zero vector, we have a zero denominator which may cause
difficulties to handle this equation: the numerator must be an
``even smaller'' zero to fulfill it.
This requirement can be fulfilled, because the third and fourth factor
of equation (\ref{eq:0mot}) simplify for $v_\mu v^\mu \longrightarrow 0$,
so that the equation takes the form
\beq
0 = v_\mu v^ \nu \eta_{\nu\rho} {d^2 x^\rho\over d\lambda^2}
  = v_\mu {1\over2} {d\over d\lambda} \left(v_\nu v^\nu\right)\ ,
\eeq
which has reduced to the {\em free} relativistic equation of motion, 
and has the solution
\beq
v_\nu v^\nu = const\ .
\eeq
This shows that
\begin{enumerate}
\item the solution $v_\nu v^\nu = 0$ is stable
  (massless particles propagate always lightlike)
\item the massless particle is apparently not influenced by the scalar 
  field $\phi$.
\end{enumerate}
The latter statement can also be seen directly from the last factors of
equation (\ref{eq:0mot}), which can be resolved to
\beq
(\eta_{\mu\nu}-{v_\mu v_\nu\over v_\kappa v^\kappa})
  {d^2 x^\nu\over d\lambda^2}
= (\delta_\mu^\nu-{v_\mu v^\nu\over v_\kappa v^\kappa})
  v_\rho v^\rho \phi_{,\nu}
\eeq
where obviously the right hand side (the 4-force) goes to zero with
$v_\nu v^\nu$.
It may even be guessed when looking again at the Lagrange function
(\ref{eq:L}), which vanishes together with $v_\nu v^\nu$, no matter
whatever the behaviour of $\phi$ is.

To provide a better understanding of the situation we encounter here, we
perform a more acurate investigation of the massless problem as a certain
limit of the massive one.  This can be accomplished in two ways:
\begin{enumerate}
\item Let the mass $m_0$ go to zero, $m_0\longrightarrow0$,
  while the energy
  \beqs
  E=p^0 c=m_0 c^2/\sqrt{1-(V/c)^2}
  \eeqs
  stays finite (``semi-constant'')
\item For fixed $m_0$ let the (3-) velocity $V$ approach $c$
\end{enumerate}

\subsection{The zero rest mass limit}

In the first case, we consider the problem with any fixed, small, finite mass
$m$, and choose the ``affine'' parameter $\lambda$ in such a way that
$v^\mu=dx^\mu/d\lambda$ is the 4-momentum $p^\mu$ of the test particle, 
i.e. 
\beq
v^\mu = {dx^\mu\over d\lambda} = p^\mu = m_0 V^\mu 
= m_0 {dx^\mu\over d\tau}
\ ,\ \
d\lambda={d\tau\over m_0}
\label{eq:lambda}
\eeq
Herewith, we have
\beq
v^\mu v_\mu
= \left({d\tau\over d\lambda}\right)^2 = {m_0}^2 = const
\eeq
so that the equation of motion (\ref{eq:eom}) reads:
\beq
{d\over d\lambda} \left({e^\phi\over m_0}{dx^\mu\over d\lambda}\right)
= m_0 e^\phi \phi^{,\mu}
\eeq
or, multiplied by $m_0$:
\beq
{d\over d\lambda} \left( e^\phi {dx^\mu\over d\lambda} \right)
= {d\over d\lambda} \left( e^\phi p^\mu \right)= {m_0}^2 e^\phi \phi^{,\mu}\ .
\label{eq:eom1}
\eeq
Going now into the massless limit, i.e.\ $m_0\longrightarrow0$, this equation
approaches
\beq
{d\over d\lambda} \left( e^\phi {dx^\mu\over d\lambda} \right)
= {d\over d\lambda} \left( e^\phi p^\mu \right) \stackrel{0}{=} 0
\eeq
where ``$\stackrel{0}{=}$" indicates equality in the limit
$m_0\longrightarrow0$\@.
Thus, in our limit, the product of the 4-momentum $p^\mu$ of the particle
with the exponential function $e^\phi$ of the potential $\phi$ at the
particle's location becomes a constant of motion:
\beq
e^\phi p^\mu \stackrel{0}{=} P^\mu = const = \left.p^\mu\right|_{\phi=0}
\ \ ,\ \
p^\mu \stackrel{0}{=} P^\mu e^{-\phi}
\label{eq:p}
\eeq
Investigating now the zeroth component of the 4-momentum, the energy
$E=p^0$, we have
\beqs
E=p^0\stackrel{0}{=}P^0 e^{-\phi}
\eeqs
from which we can obtain an expression for $e^\phi$:
\beq
e^{\phi} \stackrel{0}{=} {P^0\over p^0}={P^0\over E}\ ,
\eeq
which may be inserted into the equations for the spatial components
$p^i,\ i=1,\ldots,3$:
\beq
p^i \stackrel{0}{=} P^i e^{-\phi}={E P^i\over P^0}
\eeq
Separating the direction $n^i$ of the 3-momentum $p^i$ from its length 
$p$ according to (summation over $i=1\ldots3$ implied here)
\beq
p^i = p n^i\ ,\ \ p=\sqrt{p^i p^i}\ ,\ \ n^i n^i=1\ ,
\eeq
the original equation (\ref{eq:p}) takes the form
\beq
p n^i \stackrel{0}{=} P^i e^{-\phi}
\eeq
which enforces on $P^i$ a decomposition
\beq
P^i = P n^i\ ,\ \ P=\sqrt{P^i P^i}
\eeq
with {\em the same\/} $n^i$ as above, and the relation for the absolute
values
\beq
p \stackrel{0}{=} P e^{-\phi} = {EP\over P^0}
\eeq
Thus we have shown that along with the $P^i$, the direction vector components
$n^i$ of the 3-momentum (and thus velocity) are {\em constants of motion\/}
in the zero restmass limit,
so that the flight direction is not influenced by the scalar field,
no matter which field or particle configuration is assumed.

\subsection{The ultra-relativistic ($V\longrightarrow c$) limit}

Leaving $m_0$ now fixed, we have the equation of motion (\ref{eq:eom}),
and can again choose the parameter $\lambda$ so that
\beqs
d\lambda = {d\tau\over m_0}\ ,\ \ v^\mu=p^\mu
\eeqs
as above (\ref{eq:lambda}). The resulting equation of motion is again
(\ref{eq:eom1}):
\beq
{d\over d\lambda} \left( e^\phi {dx^\mu\over d\lambda} \right)
= {m_0}^2 e^\phi \phi^{,\mu}\ ,
\label{eq:eom2}
\eeq
which has again only three independent and one trivial component
(this can be seen by contracting with $e^\phi p_\mu$ and using
$p^\mu p_\mu={m_0}^2$).

Equation (\ref{eq:eom2}) can be evaluated to
\beq
{d\over d\lambda} p^\mu =
  \left({m_0}^2 \eta^{\mu\nu} - p^\mu p^\nu\right) \phi_{,\nu}
  = {m_0}^2 \phi^{,\mu} - p^\mu {d\over d\lambda}\phi
\label{eq:mot3}
\eeq

For the general case of the scalar field $\phi$, it is at this stage too
complicated to investigate the influence on the massive particle's
trajectory.  However, this is possible for the interesting special case of
a {\em static\/} scalar field, i.e.\ $\partial_t\phi=\phi_{,0}=0$; then
we have from (\ref{eq:mot3}) for the zeroth component:
\beq
{d\over d\lambda} p^0 = - p^0 {d\over d\lambda} \phi
\eeq
or
\beq
0 = {1\over p^0} {d\over d\lambda} p^0 + {d\over d\lambda} \phi
= {d\over d\lambda} \left(\ln\left(p^0\right) + \phi\right)\ .
\eeq
This equation has the constant integral (the ``$m_0 c$'' denominator by
choice)
\beq
\ln\left({p^0\over m_0 c}\right) + \phi = \cale = const
\eeq
so that
\beq
p^0 = m_0 c\cdot e^{\cale-\phi}\ .
\eeq
This is the energy conservation law here in its exact form.

As the zeroth component of the 4-momentum $p^0$ is generally given by
\beq
p^0 = {m_0 c\over\sqrt{1-\left(V/c\right)^2}}\ ,
\eeq
this is equivalent to
\beq
\sqrt{1-\left(V/c\right)^2} = e^{\phi-\cale}\ \ ,\ \
V = c \sqrt{1-e^{2(\phi-\cale)}}
\label{eq:encon_v}
\eeq
The energy conservation law can thus be rewritten as
\beq
V^2 = c^2 \left(1-e^{2(\phi-\cale)}\right)
\label{eq:encon1}
\eeq

\noindent
As $(V/c)^2$ must take values in the real interval $[0,1]$ only, the
difference $\phi-\cale$ must be negative, i.e.\ in the field-free ($\phi=0$)
case, the energy constant $\cale$ is always positive.  It takes the values
\beas
\cale = 0 & \longleftrightarrow & V = 0 \\
\cale = \infty & \longleftrightarrow & V = c
\eeas
In the non-relativistic limit, the constant $\cale$ for the field free case
corresponds to
\bea
\cale &=& - {1\over2} \ln\left(1-V^2/c^2\right)
\nonumber \\
&\approx& {1\over 2} {V^2\over c^2}\ \ =\ \ {E_{kin}\over m_0 c^2}\ ,
\label{eq:cale}
\eea
i.e.\ the quotient of the kinetic energy by the rest energy $m_0 c^2$
(in case of a non-vanishing scalar field $\phi$, this field occurs here
as a potential energy divided by the rest energy).

To discuss the influence on trajectories a little deeper, we investigate
now the case of a spherically symmetric, static field.  As usual, we have
motion in a plain (say the equatorial one) and angular momentum conservation,
in addition to the energy conservation from above.  From $\phi=\phi(r)$ we
have now $\phi_{,\mu}=\phi' r_{,\mu}$, and therefore for the tangential
component of equ.\ (\ref{eq:mot3}):
\beq
{d\over d\lambda} p^\varphi + {1\over r} {dr\over d\lambda} p^\varphi
= - p^\varphi {d\over d\lambda} \phi
\eeq
and thus
\beq
0 = {1\over r p^\varphi} {d\over d\lambda} \left(r p^\varphi\right)
    + {d\over d\lambda} \phi
  = {d\over d\lambda} \left( \ln\left(r p^\varphi\right) + \phi \right)
\eeq
This equation has the constant integral corresponding to the conserved
angular momentum:
\beq
\ln\left(r p^\varphi\right) + \phi = \ln\calf = const
\eeq
or
\beq
{r p^\varphi\over\calf} = e^{-\phi}
\eeq
As the $\varphi$-component of the 4-momentum is given by
\beq
p^\varphi = {m_0 r \over \sqrt{1-(V/c)^2} }{d\varphi\over dt}\ ,
\eeq
the angular momentum law reads more explicitely
\beq
{m_0\over\calf} r^2 {d\varphi\over dt} = \sqrt{1-(V/c)^2} e^{-\phi}
\eeq
or by using the energy conservation law (\ref{eq:encon_v}):
\bea
{m_0\over\calf} r^2 {d\varphi\over dt}
&=& e^{-\cale}\ \ =\ \ const
\nonumber \\
r^2 {d\varphi\over dt}
&=& {\calf\over m_0}  e^{-\cale}\ \ =:\ \ \tilde\calf\ \ =\ \ const
\label{eq:am1}
\eea
This equation does no more depend on the scalar field $\phi$ but only on
the energy constant $\cale$ and the new constant $\calf$.  If summarized
to a new constant $\tilde\calf$, this new constant resembles essentially
the specific angular momentum of Newtonian mechanics.

With these conservation laws on hand, one can eventually 
solve the equation for the trajectory approximatively.  Inserting the
usual expression for $V^2$ for motion in a plain,
\beq
V^2 = \left({dr\over dt}\right)^2  
      + r^2 \left({d\varphi\over dt}\right)^2 
\eeq
and the angular momentum law (\ref{eq:am1}) into the energy conservation
law (\ref{eq:encon1}) one obtains
\beq
V^2 = \left({dr\over dt}\right)^2
      + {\left(\calf/m_0\right)^2e^{-2\cale}\over r^2}
    = c^2 \left(1 - e^{2\phi-2\cale}\right)
\label{eq:v^2}
\eeq
or
\beq
\left({dr\over dt}\right)^2 =
  {\left(\calf/m_0\right)^2e^{-2\cale}\over r^2}
  - c^2 \left(1 - e^{2\phi-2\cale}\right)
\label{eq:dr/dt}
\eeq
Equation (\ref{eq:v^2}) has the side result that for vanishing $\phi$, 
which can be assumed at spatial infinity for a field which is localized 
anyhow (e.g., the field of a localized source), the velocity takes a 
limiting value $v_\infty$, the excess velocity, which is given by the 
relation (compare eq.\ (\ref{eq:encon1}))
\beq
{v_\infty}^2 
  = c^2 \left(1-e^{2\phi_\infty-2\cale}\right)
  = c^2 \left(1-e^{-2\cale}\right)
\eeq
(since $\phi_\infty$ must be constant, it may be absorbed in $\cale$,
and thus set to zero).

In order to obtain the trajectory instead of the time dependent motion,
the differential $dt$ in equation (\ref{eq:dr/dt}) can be substituted by
$d\varphi$ via the angular momentum law, which yields
\beq
{dr\over d\varphi} = {dr/dt\over d\varphi/dt}
  = r^2 {e^\cale\over\calf/m_0 c} {dr\over dt}
\eeq
Substituting $r$ by $u=\left(\calf/m_0 c\right)\cdot 1/r$,
equation (\ref{eq:dr/dt}) takes the form
\beq
\left({du\over d\varphi}\right)^2 + u^2 = e^{{2\cale}}-e^{{2\phi}}\ .
\label{eq:du/dphi}
\eeq
This equation of motion can be evaluated if the scalar field $\phi$ is 
specified. For large distances, it is convenient to expand $\phi$ in 
orders of $1/r$ or $u$, i.e.
(with positive coefficients $\alpha_1$, $\alpha_2$ for the attractive case)%
\footnote{
  Having in mind a model for gravity, one could e.g. set
  $\phi=-\left(A/r+\beta A^2/r^2\right)$, $A \simeq \G M/c^2$.
  Then one has $\alpha_1=m_0 c A/\calf = c A e^{-\cale}/\tilde\calf$,
  $\alpha_2 = \beta {\alpha_1}^2$.
  \label{fn:g-model}
}:
\bea
\phi &=& -\left( \alpha_1 u + \alpha_2 u^ 2 + \cdots\right)\ ,\\
e^{2\phi} &=& 1 - 2 \alpha_1 u
  + 2 \left({\alpha_1}^2 - \alpha_2 \right) u^ 2 + \cdots
\eea
Inserting this relation in (\ref{eq:du/dphi}), this equation reads, 
expanded up to the order $u^2$:
\beq
\left({du\over d\varphi}\right)^2 = 
  \left(e^{{2\cale}}-1\right) + 2 \alpha_1 u
  - \left(1 + 2 {\alpha_1}^2 - 2 \alpha_2 \right) u^ 2 + \calo(u^3)
\eeq
This trajectory equation is solved by the following trajectory, up to 
the order $u^2$:
\beq
u = u_0 \left[ 
          1 + \epsilon\cos\left\{a\left(\varphi-\varphi_0\right)\right\}
        \right]
\eeq
with the arbitrary integration constant $\varphi_0$ (which determines 
the periapsis angle) and
\bea
u_0 &=& {\alpha_1\over 1+2{\alpha_1}^2-2\alpha_2}\\
\epsilon &=& 
  \sqrt{
    1 + {1\over{\alpha_1}^2}\left(e^{2\cale}-1\right)
        \left(1 + 2 {\alpha_1}^2 - 2 \alpha_2 \right)
       }
\label{eq:eps}
\\
a &=& \sqrt{1+2{\alpha_1}^2-2\alpha_2}
\eea
($u_0$ determines the periapsis, or size of the trajectory, $\epsilon$
is the excentricity and determines its shape, while $a$ describes the
rotation of the trajectory).
The proper trajectory equation eventually reads, with these quantities:
\beq
r = {r_0\left(1+\epsilon\right) \over
  1 + \epsilon\cos\left\{a\left(\varphi-\varphi_0\right)\right\} }
\eeq
with the periapsis $r_0 = \calf/[m_0 c u_0(1+\epsilon)]$.
The expression for $\epsilon$, equation (\ref{eq:eps}), can be reformulated
to show that it approaches infinity for increasing $\cale$, corresponding
to $V\longrightarrow c$:
\bea
\epsilon &=& e^{\cale}
  \sqrt{ 2\left[
         \left(1 + {1/2 - \alpha_2 \over{\alpha_1}^2} \right)
         - e^{-2\cale}
         \left({1\over 2} + {1/2 - \alpha_2 \over{\alpha_1}^2} \right)
       \right] }
\\
&\stackrel{\cale\rightarrow\infty}{\longrightarrow}&
  e^{\cale}
  \sqrt{ 2 \left(1 + {1/2 - \alpha_2 \over{\alpha_1}^2} \right) }\ \
  = e^\cale {a\over\alpha_1}\ \
  \longrightarrow \infty
\eea
This means that the trajectory's form approaches a straight line with
increasing $\cale$, or as $V$ approaches $c$.

It would be preposterous to conclude from this fact that there is no
deflection of the fast massive particle.
The asymptotic straight lines for $r\longrightarrow\infty$ are to be
calculated as follows:
The trajectory goes to asymptotic
infinity at the poles of the denominator, i.e.\ at
$\cos\{a(\varphi-\varphi_0)\}= - 1/\epsilon$, or
\beq
\delta\varphi =
\left|\varphi-\varphi_0\right|
= {1\over a} \arccos\left( - {1\over\epsilon} \right)
\eeq
The deflection angle $\Theta$ is related to $\delta\varphi$ by
\beq
\Theta = 2 \delta\varphi - \pi
\eeq
thus
\bea
\Theta &=& {2\over a} \arccos \left( - {1\over\epsilon} \right) - \pi
\nonumber
\\
&=& {2\over a} \left({\pi\over2} + \arcsin {1\over\epsilon} \right) - \pi
\nonumber
\\
&=& \pi \left( {1\over a} - 1 \right)
\\
&&\mbox{}
 + {2\over a} \arcsin\left[e^{-\cale}/
    \sqrt{2\left\{\left(1+{\alpha_2+1/2\over{\alpha_1}^2}\right)
     - e^{-2\cale}\left({1\over 2}
        + {\alpha_2+1/2\over{\alpha_1}^2}\right) \right\} }\right]
\nonumber
\eea
For the large $\cale$ considered, the argument of the $\arcsin$ function
is small, and thus the function $\arcsin x$ is well approximated by $x$.
Then the deflection angle goes to
\beq
\Theta \approx \pi \left( {1\over a} - 1 \right)
 + {e^{-\cale} \over a}
    \sqrt{2\over\left(1+{\alpha_2+1/2\over{\alpha_1}^2}\right)
     - e^{-2\cale}\left({1\over 2}
        + {\alpha_2+1/2\over{\alpha_1}^2}\right) }
\approx \pi \left( {1\over a} - 1 \right) + {2 \alpha_1 e^{-\cale} \over a^2}
\eeq
and approaches%
\footnote{
  Using the model sketched in footnote \ref{fn:g-model}, the limiting
  deflection angle evaluates to
  \beqs
  \Theta_c = \pi\left(1-1/\sqrt{1+2{\alpha_1}^2\left(1-\beta\right)}\right)
  \approx \pi {\alpha_1}^2 \left(1-\beta\right)
  = \pi \left(1-\beta\right) \left(m_0 c A/\calf\right)^2\ .
  \eeqs
}
\beq
\Theta_c = \pi \left( {1\over a} - 1 \right)
\eeq
for large values of $\cale$, corresponding to the 
limit $V\longrightarrow c$ (compare eq.\ (\ref{eq:cale})).
This nonvanishing deflection is the result a preceding straight trajectory
caused by a trajectory ``dragging'' of the scalar interaction for
massive particles, which is essentially the same effect which can be
observed as the periapsis shift for bound elliptical trajectories.





\end{document}